\documentstyle[12pt,a4,cite,epsf]{article}
\newcommand{\be}{\begin{equation}}
\newcommand{\ee}{\end{equation}}
\newcommand{\bea}{\begin{eqnarray}}
\newcommand{\eea}{\end{eqnarray}}
\newcommand{\noi}{\noindent}
\newcommand{\nn}{\nonumber}

\newcommand{\cL}{{\cal L}}
\newcommand{\cO}{{\cal O}}
\newcommand{\cH}{{\cal H}}
\newcommand{\cK}{{\cal K}}
\newcommand{\tr}{\mbox{\rm tr}}

\begin{document}

\begin{titlepage}
\begin{flushright} hep-ph/9511465 \\CPT-95/P.3265\\ FTUV/95-56 \\ IFIC/95-58
\end{flushright}
\vspace{1cm}
\begin{center}

{\huge \bf Weak $K$--Amplitudes in the\\ Chiral and
$1/N_c$--Expansions}\\[1.5cm]

{\large {\bf Antonio Pich}$^{a}$ and {\bf Eduardo de
Rafael}$^{b}$}\\[0.5cm]
$^a$ Departament de F\'{\i}sica T\'eorica
and Institut de F\'{\i}sica Corpuscular,\\
Universitat de Val\`{e}ncia -- CSIC, E-46100 Burjassot, Val\`{e}ncia,
Spain\\[0.3cm] and\\[0.3cm]
${}^b$  Centre  de Physique Th\'eorique\\
       CNRS-Luminy, Case 907,   %\\
    F-13288 Marseille Cedex 9, France\\

\end{center}
\vspace*{1.5cm}
\begin{abstract}

It is shown that there exist symmetry constraints for
non--leptonic weak amplitudes which emerge when the
$1/N_c$--expansion restricted to the leading and next--to--leading
approximations
only is systematically combined with $\chi$PT limited to the lowest
non--trivial order. We discuss these constraints
for the couplings
$\mbox{\bf g}_8$ and  $\mbox{\bf g}_{27}$ of $\Delta S=1$ transitions and the
$\mbox{\bf B}_{K}$--parameter of $K^0-{\bar K}^0$ mixing.

\end{abstract}
\vfill
\begin{flushleft} November 1995\\
\end{flushleft} \end{titlepage}

%%%%%%%%%%%%%%%%%%%%%%%%%%%%%%%%%%%%%%%%%%%%%%%%%%%%%%%
{\bf 1.} Chiral perturbation theory ($\chi$PT) provides a useful framework to
study $K$--decays in the Standard Model.
To lowest order in the chiral expansion:
${\cal O}(p^2)$, there are only two coupling constants $\mbox{\bf g}_8$ and
$\mbox{\bf g}_{27}$ which govern non--leptonic $K$--decays. Once these
couplings are fixed phenomenologically from $K\rightarrow \pi\pi$ decays, there
follow a wealth of predictions for $K\rightarrow \pi\pi\pi$ and some other
radiative
$K$--decays\footnote{For recent reviews see e.g.
refs.\cite{Pich95,deR95,Ecker95}}.
Unfortunately, the study of chiral corrections
to these  lowest order predictions generally brings many new local
terms in the chiral effective Lagrangian of
${\cal O}(p^4)$ \cite{KMW:90,EC:90,EF:91,EKW:93}.
The number of terms is too large to make a systematic
phenomenological determination of the new couplings similar to what has been
done in the purely strong interaction sector \cite{GL85}.
The problem here is not
the complexity of the calculations; it is simply that the experimental
information we have on $\Delta S=1$ and $\Delta S=2$ transitions is too limited
when compared to the large number of
possible  ${\cal O}(p^4)$ couplings. Except
for a few remarkable predictions which have been made, one is
obliged in most cases to resort to chiral
power counting arguments and/or models
in order to make numerical estimates.

Ideally, one would like to develop
well controlled approximation methods
starting at the level of the Standard Model
Lagrangian. In that sense the $1/N_c$--expansion in QCD \cite{'tH74,Wi79},
where
$N_c\rightarrow \infty$ with
$\alpha_{s}\times N_c$ fixed, is a good candidate. Keeping only
the leading contributions in the
large--$N_c$ limit for non--leptonic $K$--decays
is however a bad approximation because, in that limit, many of the four--quark
operators of the effective Lagrangian which emerge after integrating out the
fields of heavy particles in the presence of gluon interactions are suppressed.
One has to go to the next--to--leading order
in the $1/N_c$--expansion before the
complete set of possible  four--quark operators appears.

The purpose of this note is to show that there exist symmetry
constraints for non--leptonic weak amplitudes which emerge when the
$1/N_c$--expansion, restricted to the leading and next--to--leading
approximations only, is systematically combined with $\chi$PT at the
lowest non--trivial order. Here we shall limit ourselves to spell out
these constraints for the couplings
$\mbox{\bf g}_8$ and  $\mbox{\bf g}_{27}$ and the $\mbox{\bf
B}_{K}$--parameter of $K^0-{\bar K}^0$ mixing and to the discussion of their
phenomenological implications. There are other interesting
applications of the same type for other
processes; in particular for the decay
$K_1^0\rightarrow \pi^0 e^+ e^-$ which
at this approximation can be calculated in
terms of known physical parameters, and which of course has interesting
implications for
$K_L^0\rightarrow \pi^0 e^+ e^-$ and the possibility of observing direct
CP--violation in this process. These other applications will be discussed
elsewhere.

%%%%%%%%%%%%%%%%%%%%%%%%%%%%%%%%%%%%%%%%%%%%%%%%%%%%%%%
\vspace{7 mm}{\bf 2.}
In the conventional formulation of $\chi$PT the octet of
low--lying pseudoscalar states $(\pi, K, \eta)$ are the Nambu--Goldstone bosons
associated to the ``broken'' axial generators of chiral--$SU(3)$. The
Nambu--Goldstone fields are collected
in a unitary $3\times 3$ matrix $U(x)$ with
$\det U=1$, which under $SU(3)_L \otimes SU(3)_R$
transformations $(V_L,V_R)$ transforms
linearly: $U\rightarrow V_R U V_{L}^{\dag}$.
In order to describe  non--leptonic
weak interactions it is useful to
introduce the $3\times 3$ flavour matrix vector
field

\be \label{eq:leftc}
\cL_{\mu}(x)\equiv -i\frac{f^2}{2} U(x)^{\dag}D_{\mu}U(x),
\ee
\noi
where $D_{\mu}$ denotes the covariant derivative in the presence of external
$SU(3)_{L}$ and $SU(3)_{R}$ gauge field
sources, and $f$ the $f_{\pi}$--coupling
in the chiral limit ($f\simeq 86$ MeV). Under chiral--$SU(3)$ transformations:
$\cL_{\mu}\rightarrow V_{L}\cL_{\mu}V_{L}^{\dag}$.
In terms of $\cL_{\mu}$, and
to lowest order in the chiral expansion, the operators with the same chiral
transformation properties as those of the effective four--quark Lagrangian can
then be readily obtained. They are:

\be \label{eq:lag8}
\cL_8(x)=\sum_{i}(\cL_{\mu})_{2i}(\cL^{\mu})_{i3}\,;
\ee
\noi
and

\be \label{eq:lag27}
\cL_{27}(x)=\frac{2}{3}(\cL_{\mu})_{21}(\cL^{\mu})_{13}+
    (\cL_{\mu})_{23}(\cL^{\mu})_{11}\,,
\ee
\noi
which transform respectively like $(8_{L}, 1_{R})$ and $(27_{L}, 1_{R})$
under $SU(3)_{L}\otimes SU(3)_{R}$.  To lowest order in the chiral
expansion, the effective Lagrangian of the Standard Model which describes
$\Delta S=1$ transitions between pseudoscalar states has then the
following form:

\be\label{eq:conventional}
\cL_{\rm eff}^{\Delta S=1}=
-\frac{G_F}{\sqrt{2}}\,V_{\rm ud}^{\phantom{\ast}} V_{\rm us}^{\ast}\,
\left [\mbox{\bf g}_8\,  \cL_8 + \mbox{\bf g}_{27}\, \cL_{27}\right ]+
\mbox{\rm h.c.}\, ,
\ee
\noi
with $\mbox{\bf g}_8$ and $\mbox{\bf g}_{27}$ coupling constants which are not
fixed by chiral symmetry arguments. The phenomenological determination of these
couplings  from
$K\rightarrow \pi\pi$, to lowest order in the chiral expansion,
gives\footnote{Notice that $\cL_{27}$ generates both $\Delta I=1/2$ and $\Delta
I=3/2$ transitions.}  \cite{PGR:86}

\be\label{eq:couplings}
|\mbox{\bf g}_8+\frac{1}{9}\mbox{\bf g}_{27}|\simeq
5.1\, ,
\qquad |\mbox{\bf g}_{27}|\simeq 0.29\, .
\ee
\noi
The decays $K\rightarrow \pi\pi$ and $K\rightarrow \pi\pi\pi$ have also
been  analyzed in the presence of chiral $\cO (p^4)$ corrections \cite{KMW91}.
The fitted value for  $\mbox{\bf g}_8$  decreases then by $30\%$ to $\mbox{\bf
g}_8\simeq 3.6$  while $\mbox{\bf g}_{27}$ is only slightly modified.

It is useful to go one step backwards in the theory and to analyze the
combinatorics which in the Standard Model  leads to the effective Lagrangian
above.  With one virtual $W$--field emitted and reabsorbed, and to lowest order
in the chiral expansion, there are three possible chiral invariant
configurations which give rise to the effective Lagrangian

\bea\label{eq:functional}
\cL_{\rm eff} &\!\!\!= \!\!\!& -\frac{G_F}{\sqrt {2}}\, 4\,
% \nn \\ & &
\left [\mbox{\bf a}\,\tr (Q_{L}^{(-)}\cL_{\mu})
\tr (Q_{L}^{(+)}\cL^{\mu})+\mbox{\bf b}\, \tr
(Q_{L}^{(-)}\cL_{\mu}Q_{L}^{(+)}\cL^{\mu})
\right.\nn \\ && \left.\qquad\quad\; \mbox{} +
\mbox{\bf c}\,\tr (Q_{L}^{(-)}Q_{L}^{(+)}\cL_{\mu}\cL^{\mu})\right ] \, ,
\eea
\noi
where $Q_{L}^{(\pm)}$ are the flavour matrices

\bea
Q_{L}^{(-)}=\left (
          \begin{array}{ccc}
           0 & 0 & 0 \\
           V_{\rm ud}^{\star} & 0 & 0 \\
           V_{\rm us}^{\star} & 0 & 0
           \end{array}
          \right )\qquad & \mbox{\rm and} & \qquad
Q_{L}^{(+)}=\left (
          \begin{array}{ccc}
           0 & V_{\rm ud}  & V_{\rm us} \\
           0 & 0 & 0 \\
           0 & 0 & 0
           \end{array}
          \right )\,,
\eea
\noi
which under chiral rotations transform like
$Q_{L}^{(\pm)}\rightarrow V_{L}Q_{L}^{(\pm)}V_{L}^{\dag}$.
The underlying functional integral over quark and gluon fields which gives
rise to the effective couplings in (\ref{eq:functional}) is represented
diagrammatically in Fig.~1. The solid lines
correspond to quark fields propagating
in a gluon background (the dots in the figure,)
which is subsequently integrated
down to the scales where the chiral Lagrangian of the Goldstone modes becomes
effective. The
$Q_{L}^{(\pm)}$--operators represent the emission and absorption of the virtual
$W$--field. The restriction to lowest order
in the chiral expansion implies that
at most two $\cL_{\mu}$ insertions are allowed.
When further restricted to $\Delta
S=1$ transitions, the effective Lagrangian in (\ref{eq:functional}) coincides
with the conventional one in (\ref{eq:conventional}) with

\be \label{eq:gcouplings}
\mbox{\bf g}_8=\frac{3}{5}(\mbox{\bf a}+\mbox{\bf b})-\mbox{\bf b}+\mbox{\bf c}
\qquad  \mbox{\rm and} \qquad \mbox{\bf g}_{27}=\frac{3}{5}(\mbox{\bf
a}+\mbox{\bf b})\,.
\ee

%%%%%%%%%%%%%%%%%%%%%%%%%%%%%%%%%%%%%%%%%%%%%%%%%%%
\vspace{7 mm}{\bf 3.}
Let us now examine the behaviour of
the coupling constants {\bf a}, {\bf b}, and
{\bf c} from the point of view of the $1/N_c$--expansion. It appears that the
configuration which leads to the
{\bf a}--type coupling in (\ref{eq:functional}) is $\cO (N_c^2)$, while those
leading to the {\bf b}-- and {\bf c}-- type couplings are non--leading
$\cO (N_c)$. To leading order in the $1/N_c$--expansion the coupling
{\bf a} can be calculated because in this limit the four--quark operators
factorize into current density operators and their chiral effective realization
is known from low--energy strong interaction physics to $\cO (p^4)$. With the
factor $f^2$, which is $\cO (N_c)$, included in the definition of
$\cL_{\mu}(x)$ in (\ref{eq:leftc}), the coupling constant {\bf a} is of $\cO
(1)$ in the $1/N_c$--expansion. The interesting observation is that the
factorization  result which emerges,
$\mbox{\bf a}=1$, can only be modified by gluonic configurations which are at
least next--to--next--to--leading order in the
$1/N_c$--expansion, as illustrated by the diagram in Fig.~2.
Colour matrices are
traceless, which implies that a minimum
of two gluons exchanged from one fermion
loop to the other are required to modify the factorization property, and this
leads to a relative correction of $\cO (1/N_c^2)$.
We then conclude that:
%to $\cO (N_c^2)$ and $\cO (N_c)$:

\be
\mbox{\bf a}=1+\cO \left( \frac{1}{N_c^2}\right)\, .
\ee

The configuration which leads to the {\bf c}--type
 coupling corresponds to the so
called penguin--like diagrams in the effective four--quark Hamiltonian
formulation. It is well known \cite{CFG86} that their contribution to the
coupling  {\bf c} to leading order in the $1/N_c$--expansion which in this case
is $\cO (1/N_c)$  can also be calculated in terms of known phenomenological
parameters, with the result\footnote{A detailed discussion of
this calculation can be found e.g. in ref.~\cite{deR95}}:

\be\label{eq:c-result}
\mbox{\bf c}=\mbox{\rm Re}C_{4}-16\,L_{5}\,\mbox{\rm Re}C_{6}(\mu^2)\,
\biggl [\frac{<\!\bar{\psi}
\psi\!>}{f_{\pi}^3}\biggr ]^2\simeq 0.3\pm 0.2\, ,
\ee
\noi
where in this expression $L_{5}$ is one of the $\cO (p^4)$ couplings of the
strong effective chiral Lagrangian, and $C_{4}$, $C_{6}$ are the Wilson
coefficients of the $Q_{4}$, $Q_{6}$ four--quark
operators in standard notation.
To
$\cO (1/N_c)$ the scale dependence in
$C_{6}$ cancels with the one in  $<\!\bar{\psi} \psi\!>$, while $C_4$ is
scale-independent (below the charm threshold).
The numerical result
in (\ref{eq:c-result}) comes from using the chiral limit value
$f_{\pi}\simeq f = 86$ MeV;
$L_{5}\simeq 1.4\times 10^{-3}$ and  $<\!\bar{\psi} \psi\!>(1 \mbox{\rm
GeV}^2)=-(0.013\pm 0.003)\,\mbox{\rm GeV}^3$ \cite{BPdeR95}.
The Wilson coefficients
have been evaluated using the perturbative QCD two--loop expressions
\cite{CFMR93,BJL93} restricted to $\cO (N_c)$ with\footnote{
%%%%%%%%%
More precisely, we have taken
$\Lambda_{\overline{\rm MS}}^{N_c\to\infty}\simeq 400\,\mbox{\rm MeV}$,
which for $N_c = N_f = 3$
corresponds to
$\Lambda_{\overline{\rm MS}}\simeq 300\,\mbox{\rm MeV}$.
%%%%%%%%%%%%
}
$\Lambda_{\overline{\rm MS}}\simeq 300\,\mbox{\rm MeV}$. The error in
(\ref{eq:c-result}) is partly due to the present error in the determination of
$<\!\bar{\psi} \psi\!>$, partly to short--distance
uncertainties in $C_{4,6}$.

The configuration which leads to the {\bf b}--type
coupling is also $\cO (N_c)$.
However, unlike the case of
the {\bf c}--coupling, we do not know at present how
to evaluate {\bf b} in a model
independent way, even its  leading  $\cO (1/N_c)$
contribution. There is however an
important correlation which appears at the order
of approximations which we are considering: the {\bf b}--type configuration
contributes with {\it opposite sign}
to $\mbox{\bf g}_8$ and $\mbox{\bf g}_{27}$.
This correlation of signs is in fact fully respected in the effective action
model calculation of ref.~\cite{PdeR91}, where it is found
that {\bf b} is negative (see also the recent work of
refs.~\cite{ABEFL95,ABFL95}). It is not quite respected in the model
of refs.~\cite{BBG87,BBG88}, inspired by the
$1/N_c$--expansion; but this is due to the fact that in their approach
some terms of higher $\cO(1/N_c^2)$ have also been included.

A qualitative picture towards the understanding
of the underlying physics begins
to emerge at this simple  level of approximations $\cO (p^2)$ and $\cO (1/N_c)$
which we are considering. With {\bf a}$=1$ and
fixing for example {\bf b} to the
value
$\mbox{\bf b}\simeq -0.52$,
which is the one which follows from the phenomenological
determination
$|\mbox{\bf g}_{27}|\simeq 0.29$,   % with $\mbox{\bf a} = 1$,
%in eq.~(\ref{eq:couplings})
implies
$\mbox{\bf g}_{8}\simeq 1.1$,
still too low compared to the phenomenological number
to be explained (which once corrected by the
enhancement already provided by the
$\cO (p^4)$ chiral corrections is
$\mbox{\bf g}_{8}\simeq 3.6$), but in the right direction.

We are now in the position to bring in as well the
discussion of the
$\mbox{\bf B}_{K}$--parameter which governs $K^0-\bar{K}^0$ mixing.
By analogy with the previous analysis of $\Delta S=1$ transitions, the
short-distance $\Delta S=2$ Hamiltonian can
be visualized as a convolution of two
$\bar s\to \bar d$ transitions,  with two virtual $W$--fields being emitted and
reabsorbed. Each transition results in a flavour matrix factor
$(Q_{32})_{ij} = \delta_{i3} \delta_{j2}$ times calculable short--distance loop
functions from the integration of the heavy fields. Therefore, the
effective
$\cO(p^2)$
$\Delta S=2$ Lagrangian has also the structure given in
eq.~(\ref{eq:functional}), but with the matrices
$Q_L^{(\pm)}$ replaced by $Q_{32}$. In this case the configuration {\bf c} is
identically zero because $(Q_{32})^2=0$ while {\bf a} and {\bf b} generate the
same structure
$(\cL_\mu)_{23} (\cL^\mu)_{23}$. Chiral symmetry
guarantees that the coefficients
{\bf a} and {\bf b} appearing in the $\Delta S=1$ and $\Delta S=2$
effective Lagrangians at $\cO(p^2)$ are the same \cite{DGH82}  (once the known
short-distance factors have been appropriately reabsorbed in the global
normalizations of
$\cL_{\rm eff}^{\Delta S=1}$ and $\cL_{\rm eff}^{\Delta S=2}$).
The expression which emerges for $\mbox{\bf B}_{K}$ to lowest
$\cO (p^2)$ in the chiral expansion , but taking into
account the chiral corrections which bring the chiral limit
$f$--coupling to the physical $f_{K}$, is then

\be
\mbox{\bf B}_{K}=\frac{3}{4}\,(\mbox{\bf a}+\mbox{\bf b})\,,
\ee
\noi
where, at the level of
approximation in the $1/N_c$--expansion that we are considering [i.e.,
$\cO(1)$ and $\cO (1/N_c)$], $\mbox{\bf a}=1$.
Using $\mbox{\bf b}=-0.52$ as before, one gets
$\mbox{\bf B}_{K}=0.36$, a
number which is compatible with the results of the effective action model
calculations of ref. \cite{PdeR91} as well as with various phenomenological QCD
sum rule determinations\cite{PR:85,PDPPR:91,BDG:88}.
Within errors, it is also compatible with the
results of the $1/N_c$--approach calculations of ref. \cite{BBG88},
but not with the most recent numerical estimates of
$\mbox{\bf B}_{K}$ obtained by the lattice QCD simulations
\cite{Sh94,Ietal93,Metal95,AO:95,soni:95}.

%%%%%%%%%%%%%%%%%%%%%%%%%%%%%%%%%%%%%%%%%%%%%%%%%%%%%%
\vspace{7 mm}{\bf 4.}
In order to get some insight into the underlying QCD dynamics we shall next
examine the short--distance behaviour of the two--point function correlators

\be
\Psi^{\Delta S=1,2}(q^2)\equiv i\int d^{4}x\, e^{iq\cdot x}\, \langle 0\vert
T\left(
\cH_{\rm eff}^{\Delta S=1,2}(x),\, \cH_{\rm eff}^{\Delta
S=1,2}(0)^{\dagger}\right)\vert 0\rangle
\ee
\noi  in perturbation theory and within the $1/N_c$--expansion. Here
$\cH_{\rm eff}^{\Delta S=1,2}(x)$ denote the standard $\Delta S=1$ or
$2$ four--quark effective Hamiltonians. The spectral functions associated to
these correlators describe in an inclusive way transitions from the vacuum to
physical states with total strangeness $S=1$ or $2$.
They have been calculated in
perturbation theory to next--to--leading logarithmic order in refs.
\cite{PdeR91,JP94}. The results of these
calculations give gluonic corrections of
rather normal size for the $(27_L,1_R)$ correlators (i.e., for
$\Delta S=2$ transitions and $\Delta S=1$
transitions with $\Delta I=3/2$) and a
big enhancement in the $(8_L,1_R)$ correlator. The enhancement disappears
completely when only the large--$N_c$ limit
component of the gluonic corrections
is retained.

To simplify the discussion to the essential point let us restrict ourselves to
the non--penguin operators $Q_{\pm}\equiv Q_{2}\pm Q_1$
and consider the spectral
functions associated with the
$C_{\pm}(\mu^2)Q_{\pm}$ terms in the $\Delta S=1$ Hamiltonian in the absence of
penguin--like contributions. The corresponding
results from ref. \cite{JP94} can
then be written as follows:

\be
\frac{1}{\pi}\mbox{\rm Im}\Psi_{\pm\pm}(t)=\theta(t)\frac{2}{45}N_c^2
(1\pm\frac{1}{N_c})\frac{t^4}{(4\pi)^6}\alpha_{s}(t)^{-2a_{\pm}}\,
C_{\pm}^{2}(M_W^2)
\left[ 1+\frac{3}{4}\frac{\alpha_{s}(t)N_c}{\pi}\cK_{\pm}\right],
\ee
\noi where $a_{\pm}=\pm\frac{9}{11N_c}\frac{1\mp 1/N_c}{1-6/11N_c}$ and

\be
\cK_{+}=1-\frac{30587}{\;3630}\frac{1}{N_c}+
\frac{164936}{\;19965}\frac{1}{N_c^2}-
\frac{51591}{14641}\frac{1}{N_c^3}+
\frac{\;440193}{322102}\frac{1}{N_c^4}+\cdots=
-\frac{\;3649}{3645},
\ee

\be
\cK_{-}=1+\frac{30587}{\;3630}\frac{1}{N_c}+
\frac{169706}{\;19965}\frac{1}{N_c^2}+
\frac{70335}{14641}\frac{1}{N_c^3}+\frac{1810209}{\;322102}\frac{1}{N_c^4}+
\cdots= +\frac{18278}{\;3645}.
\ee
\noi A very revealing pattern emerges when the coefficients $\cK_{\pm}$ of the
$\cO (\alpha_s)$ corrections are expanded
in powers of $1/N_c$ as shown above. In
the large $N_c$ limit $\cK_{-}=\cK_{+}$ and
the two spectral functions coincide.
The
$\cO (1/N_c)$ corrections to these coefficients are enormous, and modify the
spectral functions by the same amount but in opposite directions:
$\frac{1}{\pi}\mbox{\rm Im}\Psi_{--}$ gets a large enhancement while
$\frac{1}{\pi}\mbox{\rm Im}\Psi_{++}$ is
strongly suppressed. Although the higher
order
$1/N_c$--corrections are smaller than those
to next--to--leading order, they still
have an important overall numerical effect when compared to the exact results.
This is because in $\cK_{+}$, the alternating signs of the first five terms of
the series in powers of $1/N_c$ produce a
compensating effect, while in $\cK_{-}$
all the terms have the same positive sign which results in an important further
enhancement.

We propose to compare the \underline {relative} $1/N_c$--dependence of the
spectral functions $\frac{1}{\pi}\mbox{\rm Im}\Psi_{\pm\pm}$ calculated in
perturbation theory with those obtained to
lowest order in $\chi$PT in the chiral
limit \cite{PdeR87}, and in the $1/N_c$--expansion. We denote by
$\frac{1}{\pi}\mbox{\rm Im}\Psi_{8,27}$ the
spectral functions associated to the
effective chiral Lagrangians $\cL_{8,27}$ in eqs.~(\ref{eq:lag8}) and
(\ref{eq:lag27}). Then, with $\mbox{\bf g}_8=\mbox{\bf g}_8^{-}+\mbox{\bf
g}_8^{+}$, the ``equivalent'' spectral functions are:

\be
\vert \mbox{\bf g}_8^{-}\vert ^2 \mbox{\rm Im}\Psi_{8}
\sim \mbox{\rm Im}\Psi_{--},
\ee

\be
\vert \mbox{\bf g}_8^{+}\vert ^2 \mbox{\rm Im}\Psi_{8}\sim
\left(\frac{1}{5}\right)^2 \mbox{\rm Im}\Psi_{++} \qquad \mbox{\rm and} \qquad
\vert \mbox{\bf g}_{27}\vert ^2 \mbox{\rm Im}\Psi_{27}\sim
\left(\frac{6}{5}\right)^2 \mbox{\rm Im}\Psi_{++}.
\ee
\noi We find that the terms of relative $\cO(1)$
and $\cO(1/N_c)$ in both types of
spectral functions i.e., those obtained from
the effective chiral Lagrangian and
those obtained in perturbation theory, have exactly
the same correlation of signs
as the one implied by eqs.~(\ref{eq:gcouplings}) in the limit where
$\mbox{\bf a}=1$ and in the absence of penguins.

We shall use this comparison of relative $1/N_c$--dependence of spectral
functions as a way to suggest a plausible pattern of the $\cO(1/N_c)$ and
$\cO(1/N_c^2)$ contributions to the
couplings $\mbox{\bf a}$ and $\mbox{\bf b}$.
Setting

\be
\mbox{\bf a}=1+\alpha\frac{1}{N_c^2}+\cO(\frac{1}{N_c^3})\,; \qquad
\mbox{\bf b}=\beta\frac{1}{N_c}+\beta^{'}
\frac{1}{N_c^2}+\cO(\frac{1}{N_c^3})\,,
\ee
\noi results then in the following equivalence relations:

\be
\alpha\sim
\frac{9}{22} \ln{\left[{\alpha_s(t)\over\alpha_s(M_W^2)}\right]}
\left\{ 1 + \frac{30587}{4840} \frac{\alpha_s(t) N_c}{\pi} \right\}
+ \frac{81}{242} \ln^2{\left[{\alpha_s(t)\over\alpha_s(M_W^2)}\right]}
+ \frac{30257}{19360}
\frac{\alpha_s(t) N_c}{\pi}-\frac{1}{8}\,,
\ee
\be
\beta\sim -\frac{9}{11} \ln{\left[{\alpha_s(t)\over\alpha_s(M_W^2)}\right]}
-\frac{30587}{\;9680}
\frac{\alpha_s(t) N_c}{\pi}+\frac{1}{2}\,,
\ee
\be
\beta{'}\sim
-\frac{54}{121} \ln{\left[{\alpha_s(t)\over\alpha_s(M_W^2)}\right]}
-\frac{477}{10648}\frac{\alpha_s(t) N_c}{\pi}\, .
\ee
\noi We insist on the fact that these relations are not equalities. The
derivation of quantitative relations would require the use of dispersion
relations and a precise knowledge of the hadronic spectral functions at
intermediate energies which unfortunately is not available.  The relations
above only  show the type of $1/N_c$-- corrections in the effective couplings
which emerge if one assumes that the
$1/N_c$--behaviour of the short--distance correlators is a universal feature
of the full hadronic spectral function. The pattern suggested by this
comparison is nevertheless rather interesting. As expected, the term
corresponding to $\beta$ is large and negative. It also shows that the
$\cO(1/N_c^2)$ term which corresponds to $\alpha$ contributes with
positive corrections, which tend to cancel in the combination
$\mbox{\bf a}+\mbox{\bf b}$. The
$\cO(1/N_c^2)$ corrections to
$\mbox{\bf b}$ corresponding to $\beta{'}$  have a much smaller size.

%%%%%%%%%%%%%%%%%%%%%%%%%%%%%%%%%%%%%%%%%%%%%%%%%%%%%%
\vspace{7 mm}{\bf 5.} There are some conclusions we can draw from the previous
analyses:

The phenomenological result $\vert\mbox{\bf g}_{27}\vert\simeq 0.29$ can be
easily digested to lowest order in the chiral
expansion and to next--to--leading
order in the $1/N_c$--expansion. It requires a negative value for the
coupling constant
$\mbox{\bf b}$ to $\cO (1/N_c)$. This also helps to explain part of the $\Delta
I=1/2$ enhancement, but a quantitative understanding of the phenomenological
result $\vert \mbox{\bf g}_{8}\vert\simeq 3.6$ obtained with inclusion of the
$\cO (p^4)$ chiral corrections is still lacking at this level. As already
mentioned, a negative $\mbox{\bf b}$--coupling is a common result of various
model calculations. However, in order to explain {\it both}  $\vert\mbox{\bf
g}_{27}\vert\simeq 0.29$ and  $\vert
\mbox{\bf g}_{8}\vert\simeq 3.6$ one still needs sizable
higher
$\cO (1/N_c^2)$ positive contributions to the $\mbox{\bf a}$--coupling constant
which partly compensate in the sum
$\mbox{\bf a}+\mbox{\bf b}$, the large and negative $\mbox{\bf b}$--coupling
which is needed to get the $\Delta I=1/2$ enhancement. No model so far has been
produced which shows this convincingly; but it is interesting that both
requirements appear to be compatible with
the pattern of short--distance inclusive
calculations discussed above.

The early chiral symmetry prediction \cite{DGH82} $\mbox{\bf B}_{K}\sim 0.35$
appears then as a natural result within this scenario, but the discrepancy of
this prediction with the numerical estimates
of $\mbox{\bf B}_{K}$ obtained by the
lattice QCD simulations pose a serious puzzle
which requires further comments on
our part.

If one interprets the large lattice
results as chiefly due to the fact that {\bf b} is a very small negative
quantity or even a positive one then, from
the analysis above, it follows that the
bulk of the
$\Delta I=1/2$ enhancement has to come from penguin--like configurations
i.e., a large and positive value for the {\bf c} coupling constant\footnote{
Notice that such large corrections would also imply a large
$\epsilon'/\epsilon$ value.}. If that
is the case we have then to understand why here the
$1/N_c$--expansion, at its first non--trivial level, breaks down so
dramatically. The QCD perturbative calculation of the asymptotic spectral
function associated to the penguin $Q_6$--operator made in ref. \cite {PdeR91}
shows in fact little difference between the leading result and the one
including subleading terms in the $1/N_c$--expansion.
There is also another well known problem
in this case, which is that the predicted
value for
$\Delta I=3/2$ transitions comes out
too large:  $\mbox{\bf b}\ge 0$, results in
$\mbox{\bf g}_{27}\ge 0.6$ i.e., at least a factor of two bigger than the
phenomenological determination in eq.~(\ref{eq:couplings}).

If on the other hand we assume that the estimate
$\mbox{\bf b}\simeq -0.5$ is in the right ballpark, then
the large lattice results
for the $\mbox{\bf B}_{K}$--factor imply that the
chiral corrections to
$\Delta S=2$ transitions have to be as large as
$\cO (100\%)$! Where  could such an enormous
correction come from? The chiral loop
corrections to the $K^0-{\bar{K}}^0$ transition
amplitude have been evaluated by
several groups \cite{BBG88,BSW84,Br94,BP95}
and it is now known that,
once the terms which renormalize the $f$ coupling in the chiral limit to the
physical
$f_{K}$ are factorized, the rest of the corrections do not have large
chiral logarithmic terms. Possible large
corrections can then only come from the
local ${\cO (p^4)}$ terms of the $\Delta S=2$ effective chiral Lagrangian.
The model calculations of these couplings which so far have been made
\cite{Br94,BP95} give results which are still
controversial. These calculations are impressive but difficult to interpret.
For example, the results of the $\mbox{\bf B}_{K}$--factor  obtained in ref.
\cite{BP95} turn out to be too dependent on the choice of the
cut--off which in their approach is supposed to separate long-- and
short--distances contributions. Further progress in this direction is indeed
possible and hopefully will be made; but we
are not there yet.

In the mean time, it seems fair to conclude that
there is still, unfortunately, a large theoretical uncertainty in our knowledge
of the
$\mbox{\bf B}_{K}$--parameter. We do not understand the physics behind
sufficiently well as yet to restrict the  error bars to those of our favourite
calculation as it is done in
many phenomenological analyses of the unitarity triangle constraints.
It is important to keep in mind that the ultimate purpose of these  analyses
is to {\it test the Standard Model} and \underline{not} some particular QCD
estimate of a hadronic matrix element.

%%%%%%%%%%%%%%%%%%%%%%%%%%%%%%%%%%%%%%%%%%%%%%%%%%%%%%

\vspace{7 mm}
{\bf Acknowledgments:}

This work has been supported in part by the French--Spanish Cooperation
agreement HF94--212. The work of A.P. has been supported in part by
CICYT, Spain, under the grant AEN--93--0234.  E.de R. has benefited from the
``de Betancourt -- Perronet'' prize for his visits to Spain.

\newpage
%%%%%%%%%%%%%%%%%%%%%%%%%%%%%%%%%%%%%%%%%%%%%%%%%%%%%%%%%%%%%%%
%%%%%%%%%%%%%%%%%%%%%%%%%%%%%%%%%%%%%%%%%%%%%%%%%%%%%%%%%%%%%%%%

%%%%%%%%%%%%%%%%%%%%%%%%%%%%%%%%%%%%%%%%%%%%%%%%%%%%%%%%%%%%%%%
%%%%%%%%%%%%%%%%%%%%%%%%%%%%%%%%%%%%%%%%%%%%%%%%%%%%%%%%%%%%%%%
%\newpage

\vspace{4cm}

\noi {\Large \bf Figure Captions}

\vspace{7mm}
\noi Fig. 1\,: Diagrammatic representation of the three effective couplings in
eq.~(\ref{eq:functional}). The solid lines represent
quark fields propagating in a gluon background simulated by the dotted lines.
The $Q_{L}^{(\pm)}$--operators represent the emission and absorption of the
virtual
$W$--field. The restriction to lowest order
in the chiral expansion implies that
at most two $\cL_{\mu}$ insertions are allowed.

\vspace{3mm}

\noi Fig. 2\,: Two gluons exchanged from one fermion
loop to the other are at least required to modify their factorization  and
this leads to a relative correction of $\cO (1/N_c^2)$.

%%%%%%%%%%%%%%%%%%%%%%%%%%%%%%%%%%%%%%%%
%%%%%%%%%%%%%%%%%%%%%%%%%%%%%%%%%%%%%%%%
\newpage

%%%%%%%%%%%%%%%%%%%%%%%%%% FIGURE  %%%%%%%%%%%%%%%%%%%%%%%%
%
\begin{figure}[htb]
\centerline{\mbox{\epsfysize=20.0cm\epsffile{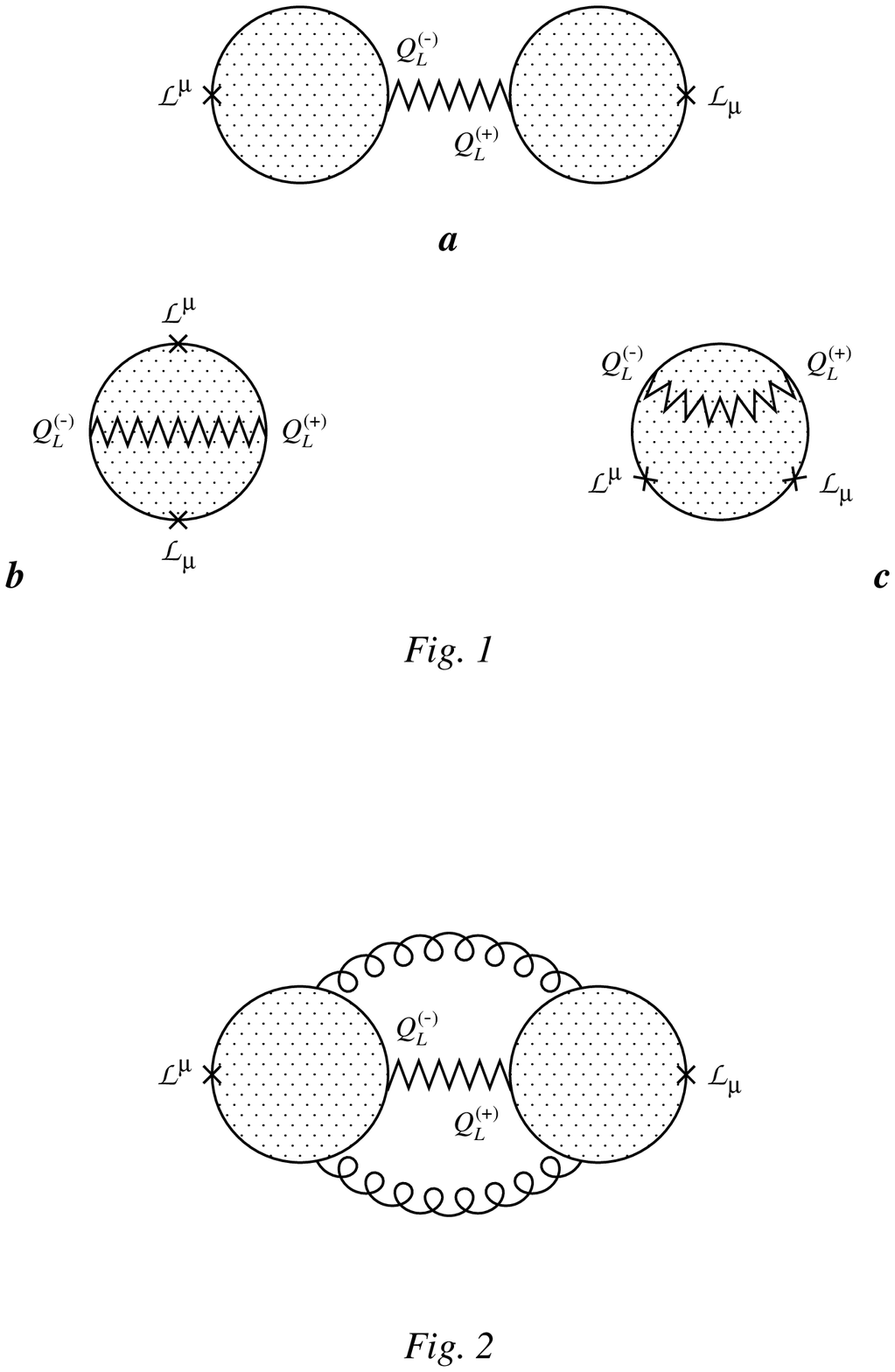}}}
\end{figure}

\end{document}